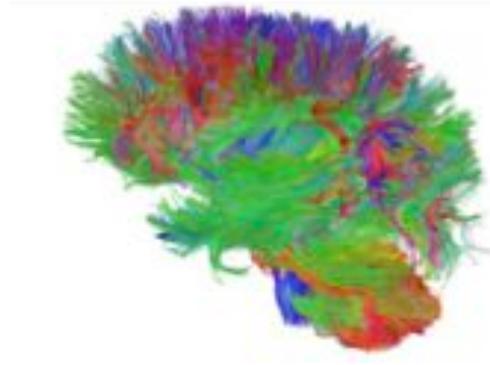

Ana M. Mihuţ, Graham Morgan, Marcus Kaiser

# Graphic Processing Unit Simulation of Axon Growth and Guidance through Cue Diffusion on Massively Parallel Processors

Technical Report No. 2
Wednesday, 14 May 2014

# Graphic Processing Unit Simulation of Axon Growth and Guidance through Cue Diffusion on Massively Parallel Processors

Ana M. Mihuţ[1], Graham Morgan[3], Marcus Kaiser[2*]

[1]Bioinformatics, School of Computing Science, Newcastle University,

[2]Neuroinformatics, School of Computing Science, and Institute of Neuroscience, Newcastle University.

[3]Game Engineering, School of Computing Science, Newcastle University.

**ABSTRACT**

**Motivation:** Neural development represents not only an exciting and complex field of study, with ongoing progress, but it also became the epicentre of neuroscience and developmental biology, as it strives to describe the underlying cellular and molecular mechanisms by which the central nervous system emerges during the various levels of embryonic development phases. The nervous system is a dynamic entity, where the genetic information plays an important role in shaping the intra- and extracellular environments, which in turn offer a reliable foundation for the stem cell precursors to divide and form neurons. Throughout the embryonic development stages, the neurons undergo different processes: migration at an immature level from the initial place in the embryo to a predefined final position, axonal differentiation and guidance of the motile growth cone towards a postsynaptic target, synaptic formation between axons and target, and lastly long-term synaptic changes which underlie learning and memory. In order to gain a better understanding of how the nervous system develops, mathematical and computational models have been created and expanded in order to bridge the gap between system-level dynamics and lower level cellular and molecular processes. This research paper aims to illustrate the potential of theoretical mathematical and computational models for analysing one important stage of neural development – axonal growth and guidance mechanisms in the presence of diffusion cues, through a visual simulation which is optimized via the graphic processing unit and parallel programming techniques.

## 1 INTRODUCTION

The research in the field of neural development is progressing at an accelerated rate, generating vast amounts of knowledge, elucidating theories through meticulous experimentation. All the different stages of neural development can be perceived as sub-systems of a whole, each of which could be described via mathematical rules and predictive computational models, in order to provide insight on how the different processes such as the molecular mechanisms responsible for the cues underlying axon guidance actually lead to the formation of the nervous system. Theoretical models have the power to represent both quantitative information in relation to a system (such as the smallest concentration gradient of a guidance cue that a migrating axon might be able to sense[1]) and the opportunity to deduce the potential consequences of the multitude of interactions that are present at molecular, cellular and network levels, therefore leading to the discovery and establishment of solid principles by which the nervous system emerges. Numerous approaches to neural modeling have been created and tailored specifically to suit a subfield or a certain type of mechanism involved at a certain stage of the nervous system development. This "modular" approach has been implied due to the large amount of input data which goes in the dynamic model, making the model therefore computationally expensive. It is quite common for computational models not only to track and accurately simulate various processes, but also to be able to learn and form predictions (through Bayesian probability networks or machine learning techniques[2]) in order to offer a deeper insight into the molecular and cellular interactions.

Axon guidance (pathfinding) refers to the ample process through which neurons allow their subsequent axons to reach out and form connection with target neurons using signaling molecules to guide the axon on the correct path. Although recent work has uncovered many of the signaling molecules that are involved in the process of axon guidance, the mechanisms underlying the phenomenon in which cells direct their movement according to gradients of chemicals in their environment (chemotaxis) are still unclear. As most of the models and simulations of axon pathfinding are computationally expensive or rely on state-of-the-art hardware (due to the large amount of neurons which have to be generated), the current paper proposes an optimized alternative, based not only on the central processing unit (CPU) power, but also on the graphics processing unit (GPU) and parallel processing techniques which greatly speed up the simulation, allowing it to be more accurate and efficient.

Even though these techniques will be applied in the context of axon guidance mechanisms (by simulating at each time step axonal growth per neuron as a parallel and independent process), they are highly reusable which suggests that they can be transferred to any other model or simulation related to the nervous system development. By being able to simulate larger and more complex neuron connectivity networks means that the applications could aid the medical field, offer insight on neurological conditions at the time of development, or even predict certain disorders such as Alzheimer's disease or epileptic seizures. The current project strives to prove that general purpose graphic processing unit implementations are suitable for scientific simulations involving particle systems, fluid dynamics, sorting and searching algorithms, random number generation, and not limited to developing large scale visual simulations. Therefore, a variety of the algorithms used for the current simulation presented in this paper were adapted and prepared to be executed on the GPU, and where hardware and the nature of the implementation permitted, the routines were parallelised using Nvidia's Compute Unified Device Architecture. Until recently, GPU programming was not widely used for scientific software development, since the GPU lacked the double or higher precision, however great advancements have been made and massive parallel processing on the GPU will most like represent the future of accelerating and simulating models for a large variety of scientific inclined fields, varying from complex physical systems to in depth molecular interactions.

---

*To whom correspondence should be addressed (m.kaiser@ieee.org).

## 2  BACKGROUND

*"The brain is a tissue. It is a complicated, intricately woven tissue, like nothing else we know of the Universe, but it is composed of cells, as any tissue is. They are, to be sure, highly specialized cells, but they function according to the laws that govern any other cells. Their electrical and chemical signals can be detected, recorded and interpreted and their chemicals can be identified; the connections that constitute the brain's woven feltwork can be mapped."[3]*

### 2.1  Fundamental principles of axonal growth and guidance through cue diffusion

**The Neuron – Core of the Nervous System:**
Neurons represent the predominant eukaryotic cell type which constructs the nervous system, along with glial cells. By means of electrical propagating signals (action potentials), the nerve cells possess the remarkable property of efficiently transferring information across vast distances. The neuron's semblance is regulated by the glial cells, which are also responsible for shaping each neuron's connectivity, providing nutritional and mechanical support.

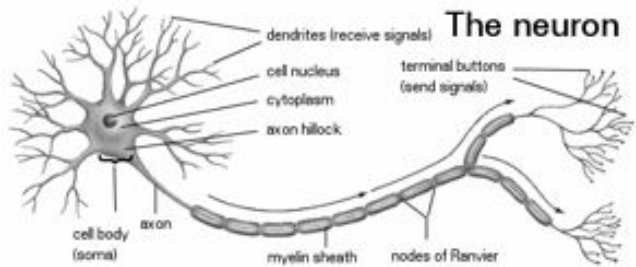

**Figure 1.** Schematic overview of a nerve cell's main components

The general structure of a nerve cell includes (Fig. 1) the soma (*perikaryon*, contains the nucleus and controls all metabolic activities, ($\emptyset soma = 20\ \mu m$)[4], the axon (nerve fibre responsible for conducting electric impulses away from the soma and towards a different neuron's dendrites), and the dendrites (tree-like structure representing a branched projection of a neuron which conducts the electromechanical stimulation to the soma). Each nerve cell possess a singular axon, ehich in turn branches out extensively and therefore transmits signals to multiple targets. The cytoplasm encapsulates numerous mitochondria responsible for converting oxygen to energy in catalysing adenosine triphosphate (ATP)[5]. The specific organelles are organized within the cytoskeleton, which consists of three different types of neurofibrils. The roles of the cytoskeleton include determining the shape of the soma, and triggering various processes which extend from the soma. The two considerable proteins involved in the nerve cell's activity are actin (which plays a vital role in axon development, specifically in the motion of growing fibres), and microtubule-associated-proteins (MAPs, which are responsible for fullfiling the anterograde and retrograde transport respectively (transporting molecules from the soma towards the axon or vice-versa). This type of transport can reach from 1mm/day up to 400mm/day depending on rapidity[6]. Each neuron has the capacity to enable synaptic connections via an ''all-or-nothing'' process known as action potential. This type of process occurs because the nerve cells maintain voltage gradients across the membrane surface (through metabolically driven ion pumps). The role of the ion pumps is to combine with the present ion channels in order to generate and maintain intracellular/extracellular concentration gradient contrasts of ions (sodium, chloride, calcium). Generally, when the voltage changes drastically, the electromechanical pulse (action potential) occurs, and manifests by traveling along the cell's axon and enabling synaptic connections which other cells upon arrival. It is important to note that any axon dysfunction is responsible for a myriad of inherited and acquired neurological disorders, which in turn affect the central and peripheral nervours systems (CNS, PNS).

**The Axon – A Rapid Form of Information Conduction:**
The fundamental role of the axon relies in its capability to transport information (undertaking the shape of electric impulses) from the soma and towards other nerve cell's dendrites, spinal cord, glands, or muscles. Axons are only about one micrometre across, but they can become extremely long. In addition to this, two types of axons can be distinguished throughout the nervous system – myelinated and unmyelinated. Myelin acts as an insulating substance, a layer of this fatty substance encapsulating the axon. In turn, myelin is synthesized by two types of glial cells (Schwann cells and oligodendrocytes). In the case of myelinated axons, gaps can be identified at evenly spaced intervals (nodes of Ranvier), giving the axon the capability to sustain saltatory conduction (rapid propagation of electric impulses). During the early neural development stages, axons grow and traverse their environment via their growth cone (situated at the very tip of the axon). The growth cone extension will always seek its synaptic target due to its dynamic actin-controlled structure. In his early research, histologist Santiago Ramón y Cajal describes the growth cone as *"a concentration of protoplasm of conical form, endowed with amoeboid movements"*[7]. The highly specialized receptors located in the growth cones recognize and respond to the various guidance cues found in the environment. Three main regions define the growth cone: a core (which contains organelles), the filopodia (a group of extensions elongating from the tip; contains receptor proteins which recognize signaling molecules). The filopodium is responsible for moving the growth cone in order to decide on which direction to extend. The process of extension is highly dependent on the interpretation of the signal coming from the surrounding molecules (attraction or repulsion can occur). Once a signal is encountered by the filopodium, the cone is stimulated and acts by advancing, retracting or turning[8]. Overall, axon elongation is the product of a process known as tip growth. In this process, new material is added at the growth cone while the remainder of the axonal cytoskeleton remains stationary. This occurs via two processes: cytoskeletal-based dynamics and mechanical tension. With cytoskeletal dynamics, microtubules polymerize into the growth cone and deliver vital components. Mechanical tension occurs when the membrane is stretched due to force generation by molecular motors in the growth cone and strong adhesions to the substrate along the axon. In general, rapidly growing growth cones are small and have a large degree of stretching, while slow moving or paused growth cones are very large and have a low degree of stretching.

**Neural development and Axon Guidance Mechanisms:**
The neuronal particularity is its wide, extended shape resulting from the vast, complex circuitry that enables it to connect and form synaptic connections with specific target cells (termination points). The main challenge of neural development is to clarify the reasons and environmental circumstances of the axons and dendrites branching, identify the correct targets, and connect with them selectively in order to create a functional neural network. As the neuronal components of the circuitry originate in different embryonic locations, the different parts of the nervous system first develop according to their own local programs like cell proliferation and cell migration[9], on the same lines as cells from other tissues of the body. Next comes the differentiation stage, unique to nerve cells, in which the global pattern of connections has to be laid down between distant components by means of axons and dendrites following specific pathways. From then on the diverse components begin to interact with each other and the last stage consists of a drastic refinement of the connections, using electrical activity of the network caused by experience of everyday life to adjust the interactions of the network.

*Axon guidance process* - The sensory function of axons is dependent on cues from the extracellular matrix which can be either attractive or repulsive, thus helping to guide the axon away from certain paths and attracting them to their proper target destinations. Attractive cues inhibit retrograde flow of the actin filaments and promote their assembly, whereas repulsive cues have the exact opposite effect. Actin stabilizing proteins are also involved and are essential for continued protrusion of filopodia and lamellipodia in the presence of attractive cues, while actin destabilizing proteins are involved in the presence of a repulsive cue. A similar process is involved with microtubules. In the presence of an attractive cue on one side of the growth cone, specific microtubules are targeted on that side by microtubule stabilizing proteins, resulting in growth cone turning in the direction of the positive stimulus. With repulsive cues, the opposite is true: microtubule stabilization is favored on the opposite side of the growth cone as the negative stimulus resulting in the growth cone turning away from the repellent. This process coupled with actin-associated processes result in the overall directed growth of an axon. Growth cone receptors detect the presence of axon guidance molecules such as Netrin, Slit, Ephrins, and Semaphorins.

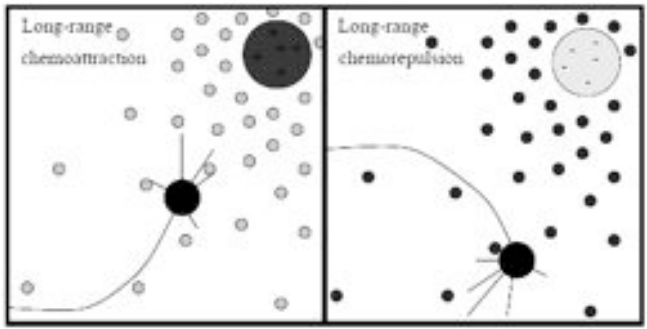

**Figure 2.** Two of the four main types of axon guidance described by R. Cajal; the long-rage chemoattraction is a process through which the target cells secrete diffusible chemoattractant substances which guide the axon from a distance; through the long-range chemorepulsion process, the axon's growth cone is repelled by diffusible factors secreted by the tissues.

It has more recently been shown that cell fate determinants such as Wnt or SHH can also act as guidance cues. Quite interestingly, the same guidance cue can act as an attractant or a repellent, depending on context. A prime example of this is Netrin-1, which signals attraction through the DCC receptor and repulsion through the Unc-5 receptor. Axon guidance directs the initial wiring of the nervous system and is also important in axonal regeneration following an injury.

**A Taxonomic Overview of Signaling Molecules:**

The growth cone structure of an axon has highly specialised receptors which identify and respond to various guidance cues found in the environment. Next, the receptors activate certain signaling molecules from the growth cone in order to decide on the cytoskeleton's new position. If the growth cone senses a gradient of guidance cue, the intracellular signaling in the growth cone happens asymmetrically, so that cytoskeletal changes happen asymmetrically and the growth cone turns toward or away from the guidance cue[10].

*Netrins* represent a family of secreted molecules (proteins) which when in range of the growth cone may either attract or repel the axon, by binding to the specialised receptors DCC and UNC5 found in the growth cone. Research supports that new axons tend to follow previously traced pathways, rather than being guided by netrins or related chemotropic factors[11]. The secondary structure presents an elevated level of conservation and it stands out that the C-terminal domain is where most of the variation between species has taken place, allowing for different amino acids to surface. These amino acids interact with specific proteins located in the extracellular matrix and these differences have led to the identification of three key netrin groups: netrin-1, netrin-3 and netrins-G[12]. Only netrin-1 plays an important role in axon guidance and the development of the central nervous system (CNS), as netrin-3 has a reduced ability to bind to the DCC receptors while netrins-G do not bind to any of the growth cone's receptors as studies in 2004 illustrate[13]. The UNC-5 receptor is mainly involved in repulsion, while DCC is a more complex receptor being able to act in both repulsive and attractive behaviour, depending on the distance from the netrin-1 source. The absence of this type of receptor causes apoptosis (process of programmed cell death – PCD), which in excess can be the direct cause of different diseases such as atrophy or cancer[14].

The attraction/repulsion mechanism functions in the presence of a protein gradient which is distributed in high concentrations at the ventral midline and gradually more diffused dorsally. Studies suggest that the presence of the gradient plays an important role in the long-range function of the UNC-6 receptor in guiding the initial axons to the midline. It was also observed that as axons reach the midline in high numbers, the temporal and spatial expression of UNC-6 becomes proportionally restricted, which suggests that the receptor is involved in specific axon guiding to more discrete locations[15].

*Slits (Sli)* – Family of proteins which are detected by the Roundabout (Robo) receptor of the growth cone. Members of the Slit family are known for being repulsive axon guidance cues. Three different types of Slits can be differentiated by their distinct domains (Slit1, Slit2, and Slit3), each containing a different number of leucine-rich repeats (LRRs), EGF repeats and a cysteine knot[16]. Slit2 binds Robo1 in a flexible linkage between the D2 domain and the first two domains of Robo1[17]. Generally, Slit interactions with the Robo1 and Robo2 receptors are the dominant element in determining whether an axon will cross the midline[18]. Robo2 in conjunction with Robo3 specify the lateral position of the axon relative to the midline. It has been recently determined that the inhibition of Robo1 receptor (colocalising with von Willebrand factor in tumor endothelial cells) can directly lead to the reduced density of micro-vessels and tumor mass of malignant melanoma. In addition to this, Robo1 has been implicated as one of the 14 different candidate genes for dyslexia[19].

*Ephrins* – Family of molecules with dual roles (bidirectional signaling) in axon guidance; generally they activate the Eph receptors which cause either attractive or repulsive reactions, however, in some isolated cases, Ephrins can also act as receptors by transducing a signal into the expressing cell (while Ephs act as ligands). Essentially, Eph and Ephrin cues respectively, control the guidance of the axons by inhibiting the survival of the axon's growth cone. The cone repels the migrating axon away from the activation site[20]. The growth cones of migrating axons do not simply respond to absolute levels of Ephs or ephrins in cells that they contact, but rather respond to relative levels of Eph and ephrin expression[21], which allows migrating axons that express either Ephs or ephrins to be directed along gradients of Eph or ephrin expressing cells towards a destination where axonal growth cone survival is no longer completely inhibited.

*Semaphorins* – Family of molecules with axonal repellent properties which are detected by two receptors, Plexins and Neuropilins. Unlike the molecules belonging to the Netrin family, the Semaphorins act as short-range inhibitory signals, deflecting the axon's growth cone from certain regions (corrects the pathway). Depending on the specific phylogenetic tree[22] and individual structure, the Sempahorins are grouped in eight major groups (all are ordered by number except for the final class, which is known as Virus or 'V'). Classes ranging from SEMA3 up to SEMA7 can be found only in vertebrates, and present a high level of versatility – e.g. SEMA3A repels axons from the facial nerves, cortical nerves, and cerebellar nerves.

*Cell Adhesion Molecules (CAMs)* – Family of protein responsible for mediating the adhesion process occurring between growing axons during the neural development stage and eliciting intracellular signaling within the growth cone.

## 2.2 Overview of existing models for axon guidance:

Recent development of computational tools has aided the field of neuroscience, by making bioelectrical activity between cells be analysed via simulations. Moreover, most of these modeling applications are capable of reconstructing the morphological structure of the nerve cells, enabling the possibility of creating vast neuronal networks[23]. Among the well-established tools in this particular field, two stand out the most[24][25] in terms of accuracy, versatility and range of functionality in neural development, as they are both built on complex, experimentally verified morphological constraints[26]. Moreover, their aim is to generate the network along with structural changes during the axonal growth and guidance processes (using intermediate time-steps rather than the final outcome). During the neural development stages, a large palette of aspects related to axon growth can be analysed and mapped in a computational model, however, due to computational efficiency, each model will only simulated one process or closely related clusters of processes. This is the case of NETMORPH and CX3D – the two applications being able to render analysed phenomena (such as dynamics of intracellular chemicals involved in axonal and dendritic outgrowth, or selection of axon growth direction following guidance cues in the environment[27]). The downside of this approach is that the statistics of the morphological changes are computed without being verified by a mathematical model of intracellular of extracellular processes which ultimately lead to those changes. Depending on the model's theoretical nature, its area of interest, and its development hierarchy, neural development models could be classified in the following ways:

**Formality**: Formal models are generally expressed through mathematical equations (generally shape the structure of a system, its starting parameters, conditions of existence), or through computer programs/ programmed routines[28], unlike informal methods which are commonly represented using dependency diagrams. Building a formal model implies that a formal language was used extensively in order to eradicate any inconsistencies or hidden assumptions and therefore reinforce the robustness and precision of the final model[29]. The main aspiration of formal models consists in their ability of letting component-component interaction surface, allowing researchers to test the plausibility of hypothetical mechanisms[30].

**Hierarchy of design**: A top-down or a bottom-up approach can be adopted when stipulating formal models. A top-down approach assumes that the model incorporates elements along with their interactions which are responsible for enabling specific model properties. The bottom-up design approach assumes that the pre-described behaviour does not exist, and therefore the interactions between the elements are investigated.

**Phenomenological models**: This type of model generally replicates the experimental data, without requiring the mathematical relationships or parameters to correspond to the underlying biological processes. At first glance, this type of model seems to lack consistency, however it can be informative and it is commonly used as a forerunner to a specifically tailored mechanistic model.

**Mechanistic models**: Generally attempt to investigate the consequences of a selected set of processes, or explore the essential aspects of the mechanisms with a tighter reference to the underlying biological and physical processes. A complete mechanistic model is extremely difficult to describe and develop and will always be described by a phenomenological model at a higher level.

*NETMORPH* – Represents a scientific application developed within VU University of Amsterdam – department of Experimental Neurophysiology, built on top of a collection of verified mathematical models describing the dendritic and axonal arbor formation and growth[31].

The application generating the simulation was developed using the C++ programming language in order to make use of its extended memory management capabilities and therefore increase rendering efficiency. It encompasses a set of template model components but it also allows the user to define new parameters for the model or simply build on top of the existing parameters. At the core of the simulation application lies the stochastic phenomenological model of Van Pelt[32] which describe the growth of neurons over time, from the perspective of individual growth cones.

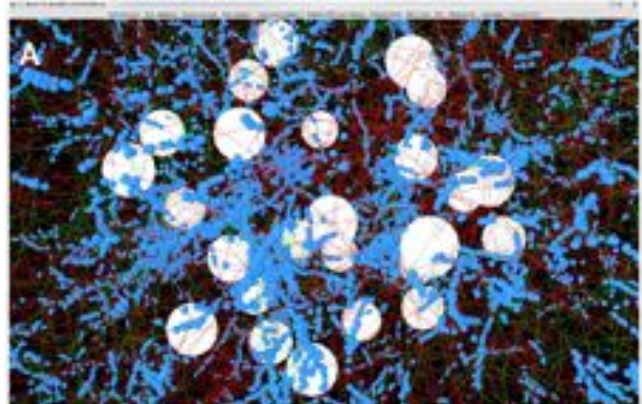

**Figure 3.** Capture of a simulation created using NETMORPH.

Being a highly technical application used for research purposes, it features only command line interaction and not a robust graphical user interface which implies that the number of commands one can run is very restrictive and the overall application does not leave room for customisation. However, it does provide a two dimensional and a separate three dimensional channel for rendering the simulation, facilitating the analysis of the different processes which take place at each time step.

*CX3D* – complex modeling application which encompasses all stages of corticogenesis (from cell division to cell-cell contact and diffusible signals)[33]. The need for such an application rose when changes in the gene expression during neural development had to be understood along with the mechanisms through which gene expression levels drive processes such as differentiation or migration.

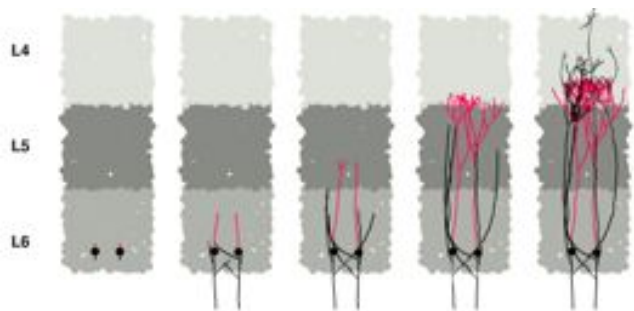

**Figure 3.** CX3D preview of axon branching overlaying gradient density

Like various scientific applications written in Java, CX3D is open-source, which means that its functionality grows with the number of users developing new content for it. Due to using a physics engine which computes the forces between elements, it is capable of accurately simulating the diffusion of substances through the extracellular space. In the figure above the branching pattern based on extracellular signaling molecules is illustrated; the different cortical tissues differentiated by the L4, L5, and L6 markers respectively are composed of different types of cells (different colour gradients). An advantage of CX3D is that it has NeuroML extensions[34] in order to use descriptions of electrophysiological nature. Due to the fact that the application relies on the Java Virtual Machine (JVM), the execution speed directly depends on the proportion of the physical dynamic model. Moreover, CX3D was not developed as a multithreaded or parallel application, because then it would be dependent on operating system version, hardware specifications (OpenGL version), or parallel processing architectures (CUDA). Even though both NETMORPH and CX3D lack the correlation between synapse formation and electrical activity, both feature extensive modeling functionality which could be further developed and optimised.

**NeuroConstruct** – represents a neural network modeling software developed by University College London in the Department of Neuroscience, Physiology and Pharmacology. It was created using Java and can manipulate script files for various platforms including Neuron, Genesis and PyNN by using the latest NeuroML specifications. Generally, the models created with NeuroConstruct incorporate dendritic morphologies and cell membrane conductase[35].

## 3 METHODS

The following simulation aims to reconstruct the process of axon growth and guidance through cue diffusion in a three dimensional environment by concomitantly making efficient use of the graphic processing unit (GPU) and parallel processing techniques. The following formal mechanistic model has strong theoretically verified mathematical foundations, which make the simulation accurate and reliable. Another important aspect of the simulation consists of its ability to export data regarding the environment at given time-steps (such as the number of connections made, the overall density, the total number of nerve cells versus computation efficiency, overall average rendering time, nerve cell and growth cone positions). This data has a statistical importance, as it enables the implementation of other analysis and predictive techniques. Two types of visual simulation will be outputted, a direct axon guidance, where the axon's motile growth cone structure is not influenced by changes in gradient, or diffusing cues, and will only form a connection if it is in range of another neuron. The secondary simulation raises the level of complexity described by the former algorithm by implementing particle systems to represent diffusing cues which play the role of chemoattractants or chemorepellents during the axon guidance process. This model does not account for the various types of substrates (inhibitory or repulsive), but new functionality could be added relatively easy to the current computational framework.

### 3.1 Description of underlying technology used

Similarly to NETMORPH, the current application was developed using the C++ programming language due to its memory management functionality and ease of implementing both the OpenGL rendering pipeline and CUDA parallel processing architecture.

*3.1.1 The OpenGL rendering pipeline* (Open Graphics Library) was originally developed by Silicon Graphics and represents the open graphics standard in the world due to its cross-language, multi-platform Application programming interface(API). Through the API interaction with the GPU, hardware-accelerated rendering is achieved. In the present, OpenGL is maintained by the non-profit consortium known as Khronos Group. The version used for developing this application is OpenGL 4.1. due to its ability to define multiple viewports and scissor rectangle which would then be used when generating several scenes at once from a geometry shader. For the purpose of this application only, the Open Computing Language (OpenCL) could have been adopted in order to enable specific parts of a program to access the GPU for non-graphical computing. However, most of the calculus and high-end computation will be implemented using the official OpenGL shading language (GLSL).

*3.1.2 GLSL* represents the high-level shading language which has its syntactic roots established in the C programming language. GLSL was created in order to provide the user with a more direct way of controlling the graphics pipeline without the need of using hardware-specific languages. Although nowadays it is used solely for graphics rendering and computing the different mathematical operations necessary for outputting geometry to the screen, it is the best way in which graphic processing can be accelerated. The non-geometrical data which does not have to be rendered to screen will be processed in parallel on the GPU cores via the CUDA architecture which is now commonly used for General Purpose Graphic Unit Processing (GPGPU computing). The reason GLSL is used in conjunction with CUDA is because the simulation makes use of both GPU and GPGPU computing and CUDA would not offer the functionality to carry both types of computations at the same optimal level. User-defined functions are supported, and a wide variety of commonly used functions are built-in as well. This offers the graphics card manufacturer the ability to optimise these built-in functions at the hardware level if they are inclined to do so. Many of these functions are similar to those found in the standard mathematics library of the C programming language. In order to link CPU side (regular C++ syntax) variables which do not change every frame to GLSL shaders, the keyword "*uniform*" must be used. This is a global GLSL variable, which does not change from one rendering call to another unlike the input and output variables which change depending on shader stage.

*3.1.3 Nvidia CUDA* Compute Unified Device Architecture enables the user to find parallel processing opportunities in the ported code, and through kernel encapsulation, the ported code can execute on multiple threads on the GPU. Unlike OpenGL (which is a specification of most graphics hardware), CUDA is strictly supported by Nvidia graphics cards. However, CUDA-capable GPUs contain hundreds of cores that can collectively run thousands of threads (as each core has shared resources such as a register file and memory). This specific architecture has the advantage that the shared memory located on the chip allows parallel tasks running on the cores to share resources without transferring them over the system memory bus. Unlike the CPU, the GPU is specialised for compute-intensive, highly parallel computation and in consequence it was designed such that more transistors are devoted to data processing rather than flow control and data caching. Because in parallel programming the same program is executed for each data element, there is a lower requirement for sophisticated flow control, and because it is executed on many data elements and has high arithmetic intensity, the memory access latency can be hidden with calculations instead of big data caches. Data-parallel processing maps data elements to parallel processing threads. Many applications that process large data sets can use a data-parallel programming model to speed up the computations. In 3D rendering[36], large sets of pixels and vertices are mapped to parallel threads[37]. At the core of CUDA lie three key abstractions – a hierarchy of thread groups, shared memories, and barrier synchronization – that are simply exposed to the programmer as a minimal set of language extensions. They guide the programmer to partition the problem into coarse sub-problems that can be solved independently in parallel by blocks of threads, and each sub-problem into finer pieces that can be solved cooperatively in parallel by all threads within the block. Each block of threads can be scheduled on any of the available multiprocessors within a GPU, in any order, concurrently or sequentially, so that a compiled CUDA program can execute on any number of multiprocessors, and only the runtime system needs to know the physical multiprocessor count.

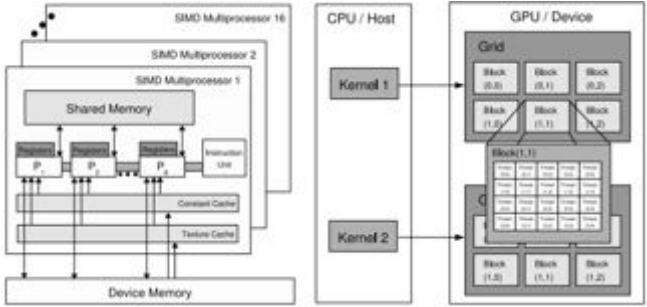

**Figure 4.** The CUDA hardware interaction interface (left hand side) presents a collection of multiprocessors, where each one contains shared memory, common to all processors inside it, 32-bit registers, cached memory and a texture. The CUDA programming model (right hand side) represents a collection of threads running in parallel. A collection of threads (warp) can run simultaneously on a multiprocessor, however the size of the warp is hardware dependent and differs from GPU to GPU.

A GPU is built around an array of Streaming Multiprocessors (SMs). A multithreaded program is partitioned into blocks of threads that execute independently from each other, so that a GPU with more multiprocessors will automatically execute the program in less time than a GPU with fewer multiprocessors. There is a limit to the number of threads per block, since all threads of a block are expected to reside on the same processor core and must share the limited memory resources of that core. On current GPUs, a thread block may contain up to 1024 threads. However, a kernel can be executed by multiple equally-shaped thread blocks, so that the total number of threads is equal to the number of threads per block times the number of blocks. Blocks are organized into a one-dimensional, two-dimensional, or three-dimensional *grid* of thread blocks. The number of thread blocks in a grid is usually dictated by the size of the data being processed or the number of processors in the system, which it can greatly exceed.

*3.1.4 Nvidia Nsight* is a new development platform which is integrated within the Microsoft Visual Studio IDE and helps facilitate the debugging process of CUDA and GPU code. Some of its main features include tracing multi-core CPU and GPU activities on a single timeline, correlating OpenGL and CUDA activities to the exact CPU code line, and specialising reports and analysis on application efficiency (from frame rate, to processing speed, memory load and general statistics of the application run). In addition to this, it provides specific functionality for OpenGL and GLSL alike, consisting in a frame debugger with render state and draw call inspection, enables debugging oh shader code directly on the GPU hardware. In the case of this simulation, Nsight was used in two distinct ways: to debug the kernels specific to CUDA, and to output performance statistics and analysis logs which will be described into detail in the results section.

## 3.2 Initial random uniform distribution algorithms

The application proposed and evaluated throughout this paper can be separated in four different phases: generating initial neurons to their fixed position, generating an axon for each neuron with a random direction and attributing the pivot element to the neuron, the direct guidance model, and the cue diffusion guidance model.

The initial phase takes as input the neuron diameter size (in units) and the total number of neurons, scales the neuron model to the specified size and uses one of the following algorithms to generate the 3D coordinates for each neuron in the system. The four models presented below are variations of the uniform random N-sphere ($S_n = \left\{x \in \mathbb{R}^{N+1} \big| \|x\|_2 = 1\right\}$) distribution algorithm which were implemented in the simulation due to their different run-time efficiency.

The uniform random sphere distribution algorithm has the unique property that no two coordinate sets collide[38], and all coordinates are evenly distributed between one another. At the core of this routine lies the Monte-Carlo approach, which generates a random point in $X \in [-1,1]^{N+1}$ interval and rejects it if it lies outside the sphere's bounds ( $\|X\| > 1$). If it resides within those bounds, $U = \|X\|^{-1} \cdot X$ gives a random point belonging to the unit sphere S (exception where X evaluates to 0), and produces a uniform distribution on the sphere. Although this approach guarantees $O(n^2)$ running time complexity and would not negatively affect small sets of points, it grows worse than exponentially in N. This is due to the volume of the unit sphere decreases more than exponentially as $N \to \infty$, where $\Gamma(\cdot)$ represents the gamma function.

$$V(N) = \frac{\pi^{\frac{N}{2}}}{\Gamma\left(1+\frac{N}{2}\right)} \quad (1)$$

$$\Gamma(x) = \int_0^\infty e^{-t} \cdot t^{x-1} dt \quad (2)$$

***Normal deviate method*** – represent the method through which the (x, y, z) coordinates are chosen from a normal distribution of mean 0 and variance 1. The coordinate vector is then normalised. Before undergoing the process of normalisation, it has a density that depends only on the distance from its origin. According to Knuth's algorithm, a random variable has distribution *N(0,1)* if it has the density function $f(x) = \frac{1}{\sqrt{2\pi}} e^{\frac{-1}{2x^2}}$.

In the case of a d-dimensional random vector X has distribution *N(0,1)* if its components are independent and have distribution *N(0,1)* each. Then, the general formula for X's density is the following: $f(x) = \frac{1}{(\sqrt{2\pi})^d} e^{\frac{-1}{2}<x,x>}$.

The latter condition follows from the Fourier transform properties and the particular form of the Fourier transform of the normal distribution[39]. If we also consider $U \in \mathbb{R}^{d \times d}$ to be an orthogonal matrix (i.e. $UU^n = U^n U = I$) then $Y = UX$ has distribution *N(0,1)*.

For $\forall$ finite set $A \in \mathbb{R}^d$ the following holds true:

$$P(Y \in A) = P(X \in U^t A) =$$
$$= \int_{U^t A} \frac{1}{(\sqrt{2\pi})^d} e^{\frac{-1}{2}<x,x>} =$$
$$= \int_A \frac{1}{(\sqrt{2\pi})^d} e^{\frac{-1}{2}<U_x,U_x>} =$$
$$= \int_A \frac{1}{(\sqrt{2\pi})^d} e^{\frac{-1}{2}<x,x>} \quad (3)$$

Every rotation function is represented as an orthogonal matrix multiplication, so the conclusion from the above relation is that normally distributed random vectors are invariant under rotation and therefore generating *X* with distribution *N(0,1)* and then projecting it onto the unit sphere produces random vectors $= \|X\|^{-1} \cdot X$ that follow a uniform distribution on the sphere (not just on its surface). By using the Box-Muller approach, the computational model for generating *X* has an overall linear complexity.

***Trigonometry method*** – is an approach that assumes the space in context is represented by a three-dimensional sphere (frequently entitled "2-sphere" due to being restricted to two degrees of freedom). The essential idea behind this technique is that the area defined by a sphere is equal to the area of any right circular cylinder circumscribed about the sphere (excluding the bases). Each of the three coordinates of a uniformly distributed point on the unit sphere is uniformly distributed on [-1,1] (but the three are not independent, obviously). Therefore, it suffices to choose one axis (Z, say) and generate a uniformly distributed value on that axis. This constrains the chosen point to lie on a circle parallel to the X-Y plane, and the obvious trigonometric method may be used to obtain the remaining coordinates. The first step in the algorithm is to select z uniformly distributed in the interval [-1,1], followed by the selection of t uniformly distributed on the interval [0, 2*pi], then defining a set of three coordinates – r (sqrt(1-z^2)), x (r*cos(t)), and y (r * sin(t)).

***Coordinate method*** – initially obtains the distribution of a single coordinate of a uniformly distributed point on the N-sphere. Then, it recursively gets the distribution of the next coordinate over (N-1)-sphere, and so on. Fortunately, for the usual 3D space(i.e. 2-sphere), the distribution of a coordinate is uniform and one can do a rejection sampling on 2D for the remaining 1-sphere(i.e. a circle). Consider the existence of a vector *U* that is uniformly distributed on the N-sphere space $S_N$, its projection $U_1$ and its coordinate x. The density function $f_{1(x)}$ of $U_1$ is defined on the interval *[-1,1]* with values in $\mathbb{R}^+$. The function $f_{1(x)}$ can be expressed via the incomplete beta function:

$$B_x(a,b) = \int_0^x t^{a-1} \cdot (1-t)^{b-1} dt, \text{ where } B(a,b) = B_1(a,b). \quad (4)$$

In order to determine $f_1$ the point *u* is selected, (where *u* belongs to the N-Sphere space) along with a direction to a new offset position $\tilde{u} \in S_N$. The covered distance between the two positions is

$$\|\tilde{u} - u\| \approx |arcsin(\tilde{x}) - arcsin(x)|$$

As $\tilde{u} \to u$, an equality can be observed

$$\frac{\partial u}{\partial x} = arcsin(x)' = \frac{1}{\sqrt{1-x^2}}$$

The N-sphere radius is $\sqrt{1-x^2}$, so the sphere's volume equals

$$R(x) = V(N-1) \cdot \left(\sqrt{1-x^2}\right)^{N-1}$$

The distribution of $U_1$ has to be linear in both *R(x)* and $\partial u / \partial x$,

$$f_1(x) = s \cdot arcsin(x)' \cdot \left(\sqrt{1-x^2}\right)^{N-1}$$
$$= B\left(\frac{1}{2}, \frac{N}{2}\right)^{-1} \cdot \left(\sqrt{1-x^2}\right)^{N-2}$$

where $x \in [-1,1]$. The scaling factor *s* assures that the integral over $f_1$ is 1, and it evaluates to the following

$$s = \left(\int_{-1}^1 \left(\sqrt{1-x^2}\right)^{N-2} dx\right)^{-1} = B\left(\frac{1}{2}, \frac{N}{2}\right)^{-1}$$

Using the density $f_1$, a series of random values can be generated by integrating $f_1$ and inverting $F_1$.

$$F_1(x) = \int_{-1}^x f_1(y) dy = \frac{1}{2} + sign(x) \cdot \frac{B_{x^2}\left(\frac{1}{2}, \frac{N}{2}\right)}{2B\left(\frac{1}{2}, \frac{N}{2}\right)}$$

Newton's method is used for calculating an approximation of $F_1^{-1}$, then a recursive approach is employed in order to generate a random point, distributed uniformly on the (N-1)-sphere. The stopping condition of the recursion step of the algorithm is when *N = 0*; at this stage, the algorithm randomly selects a point in the interval {-1, 1}. The run-time complexity of this approach is linear, however, the approximation step (where Newton's method is used), even if it is rapidly convergent, still restrains the overall performance considerably.

## 3.3 Implementation of collision detection methods

In most physically accurate simulations that involve rigid bodies, collision detection is approached as a two tiered process – the detection phase and the response phase. As the axon (guided by the growth cone) travels through the environment, it will reach the close proximity of a neuron, and if all the conditions are met (range of detection, concentration of guidance molecules), the growth cone will form a connection with the dendrites of the neuron. It is therefore important to detect a collision of this nature and simulate it appropriately. Since one of the objectives of the simulation is to render large amounts of nerve cells, the collision detection algorithm has to be efficient and this can be achieved if it works locally, on clusters of neurons, rather than on the whole model. The simulation will not need to take into consideration the second step of the collision process (collision response), as it is only concerned with the formation of neural connections. The general collision detection algorithm implemented in the simulation takes a *divide et impera* approach by detecting in which groups of objects collision is likely to happen (broadphase), and then applying the collision detection algorithm to each element in the specific group (narrowphase).

***Sphere-Sphere Collision*** – assumes that the components of the simulation can be represented as a sphere centred on the component's position vector, then determine if two spheres intersect. A collision takes place if the distance between the centres of the two spheres is less than the sum of the radii of the respective spheres. Pythagora's theorem is then used in order to calculate the distance between the spheres and compare it to the sum of the radii.

$$d = \sqrt{(x_2 - x_1)^2 + (y_2 - y_1)^2 + (z_2 - z_1)^2}$$
$$d < r_1 + r_2$$

The approach described above is straightforward, yet unrefined and not very efficient, since a square root is expensive to compute, especially on a large dataset. A premature optimization will be implemented at this stage, and instead of using a square root operation, the algorithm will compare $d^2$ and the square of the sum of the radii.

### 3.4 The "Direct Guidance" approach

The "direct guidance" method assumes that the growth cone travels on a random, predetermined direction until it either collides with the dendrites of another neuron, or until it leaves the simulation area. In the simulation created by Kaiser et al.[40], a connection was taken into consideration if the Manhattan distance between the growth cone of the axon and the dendrites of another neuron was of one unit. The collision detection algorithm used spheres to encapsulate both the neuron and the axon, making the collision detection efficient but not as accurate as possible. The "direct guidance" approach modeled in the current simulation uses sphere-sphere collision detection, however, for the cue diffusion simulation, a different approach is proposed. This stage of the simulation commences by using one of the four random uniform n-sphere point distributions discussed before (this option can be chosen via command line) in order to render each neuron to its correct location. These coordinates are stored as Vector3 types (the Vector3 type is defined by three floating-point values, which can then be normalised or used in common vector operations such as multiplication, distance, or length), in a vector data structure. Frame after frame, each Vector3 value (containing the x, y, and z coordinate respectively) will be sent as input to the vertex shader (GPU), which will compute the position of the pixels to be drawn and send the information to the fragment shader (still GPU), which is responsible for producing the final output on screen.

For each neuron in the simulation an axon direction is randomly attributed and stored in a list. The axon's initial position lies tangent to the neuron on the random direction specified. Since the mesh representing the axon takes the shape of a capsule, at each time step another capsule will be added at the end of the previous one on the direction computed until the end pivot of the capsule is in 0.05 units of another neuron, or until it is out of the simulation bounds. The axon's offset (the position at which the new part of the axon is added) is not stored, nor computed on the CPU – it is in fact computed via the vertex shader, that in turn feeds, in each frame, the new offset for each axon part). The efficiency of this approach consists in the fact that mesh data is not stored as duplicates (not for neurons nor for axons); the meshes are only defined once, and in each frame the same neuron mesh is being rendered in different positions, while concomitantly for each neuron, the axon mesh is being rendered along with the offset until the collision detection algorithm finds a hit or until the simulation's environment limits are breached. This approach will produce a final 3D environment showing all neurons and their elongated axons, and offers the user the capability of freely moving around via the camera implementation. Each connection (defined by the two neuron positions, the axon's length and its direction) will be stored. Based on the collision detection test, an adjacency list is created using CUDA. This list is then traversed to create the graphical output.

Even though the common graph operations are completed in practical times using parallel algorithms, this takes place at a high hardware cost[41]. The adjacency list along with the breadth first search (BFS) approach to graph traversal use one thread per vertex, with all threads being multiplexed on 128 processors by the CUDA environment. For the implementation of the BFS algorithm, the device shared memory is not used, as the vertex data can be present at any location in the global edge array. At each iteration, the number of vertices being processed in parallel will be expanded. The current neuronal network can be easily treated as a graph, and therefore represented as a compact adjacency list, which is then packed in a single large array. Each vertex will point to the head (beginning index) of its specific list in the edge array. For a graph $G(V,E)$, the vertices are represented as array $V_a$. Another array of the list stores the edges ($E_a$) with edges of vertex $i+1$ following the edges of vertex $i$ for all $i$ in $V$.

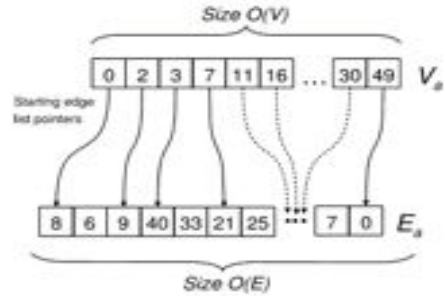

**Figure 4.** Schematic representation of the adjacency list.

***Adjacency list traversal*** – In order to traverse the list and output the neurons and their respective axons on a given direction, the BFS approach with level synchronisation is employed.
BFS method will traverse a level of the list and not iterate over it again, therefore creating a level *frontier* (corresponding to all the nodes being processed at the current level). If CUDA was not used, the normal approach while traversing the list would be to store the vertex at each level in a visited queue. However, this would slow the execution speed on the CUDA model. In order to maintain an efficient implementation, a thread is allocated to each vertex. The difference between a CPU and a CUDA approach to this routine is that in the case of the CUDA implementation, the algorithm of traversal needs to be repeated until *frontier* is empty, while the CPU side code terminates when all the levels of the graph are traversed and when frontier is empty, rendering the GPU code twice as efficient.

```
Algorithm    CUDA_BFS_KERNEL (V_a, E_a, F_a, X_a, C_a)
1: tid ← getThreadID
2: if F_a[tid] then
3:    F_a[tid] ← false, X_a[tid] ← true
4:    for all neighbors nid of tid do
5:       if NOT X_a[nid] then
6:          C_a[nid] ← C_a[tid]+1
7:          F_a[nid] ← true
8:       end if
9:    end for
10: end if
```

**Figure 5.** Pseudocode illustrating proposed adjacency list traversal

### 3.5 The "Cue Diffusion Guidance" approach:

The cue diffusion approach to axon guidance, unlike the direct approach discussed in the previous section, makes use of diffusing molecules to guide the axon's motile growth cone to its final target. This implementation will make use of the random uniform sphere distribution algorithms for positioning the neurons to their final location and avoiding collision detection checks (as described previously), and in addition to that, it will also implement the same sphere – sphere collision detection approach, however, a collision – response method will also augment the algorithm in order to determine the final orientation (angle and velocity) of the axon's growth cone after being in range of the gradient. During the development of the central nervous system it is crucial that axons reach their correct targets via guidance from molecular gradients[42,43,44] (chemoattarction/chemorepulsion mechanisms). In order to respond to a gradient cue, an axon must be able to both make a decision regarding gradient direction and then to convert that decision into a directed motile function[45]. Recent research and models have addressed the mechanisms involved in the decision-making step[46], however it is still largely unknown how the decision is subsequently converted into a change in the behavior of the growth cone. The current paper considers the immediate and biased turning mechanism, whereby the growth cone responds directly to the change detected in the nearby gradient (turns up the gradient in case of attraction or down the gradient in case of repulsion). This approach greatly simplifies the myriad of molecular processes and the various types of molecular cues involved at axon guidance level, however, it is still accurate and represents a strong foundation for further, more detailed modeling of this stage in neural development. A biased and immediate turning is commonly observed during in vitro experiments that examine the response of axons to steep gradients in two dimensions[47,48]. In the context of 3D in vitro experiments, the collective growth of a population of axons is biased by the gradient, unlike the 2D case[49]. In order for the computational model to simulate the two guidance and turning mechanisms, a discretization of a broader mathematical model had to be employed.

**Diffusion of guidance molecules** – *"A particle system is a collection of many minute particles that together represent a fuzzy object. Over a period of time, particles are generated into a system, move and change from within the system, and die from the system."*[50]. The implementation of a particle system on GPU side was relatively straightforward and could be illustrated in three different steps: integration, data structure construction, collision processing.

The initial integration step involves including the particle attributes (position and velocity) to move the particles throughout the environment. For simplicity, Euler integration was chosen, however, Lagrangian integration could easily replace the current implementation (as it requires less storage and bandwidth considering it only stores model properties at particle positions, rather than at every point in space)[51]. At each time-step, the position of a particle within the system will be updated by velocity, and its velocity will initially be sampled randomly from a set range, and over time it will be affected via applied forces. Both particle positions and velocities will be stored as Vector4 f arrays, with the positions being stored in a vertex buffer object (VBO), ready to be processed via the OpenGL pipeline. The specific VBO memory location is mapped in order to be accessed by CUDA (via "cudaGLMapBufferObject"). As OpenGL standards recommend, these arrays will be double-buffered – while a new value updates, it will not affect the particles which were not processed and rendered at the time. In this particle system it is assumed that the particles are independent from each other and do not interact in order to change trajectories over their respective lifetime. For local interactions, such as collisions, the spatial subdivision and grid data structure were employed. The force of interaction will decrement based on distance travelled from diffusing source, and therefore the force for any particle can be computed by comparing it to its neighbours within the particle's radius. A uniform "loose" grid was chosen to handle the neighbour subdivision[52] (which can be easily replaced by more efficient or sophisticated structures such as hierarchical lists). The assumption made before implementation is that each cell in the matrix coincides with the physical size of the particle ($2r_{particle}$). In this manner, one particle can only cover a maximum of eight cells in a three dimension coordinate system. During the particle collision detection phase, the particles in neighbouring cells (a total of 27 at a time) have to be examined, since a particle can overlap several cells at a time. This approach is optimal, since it does not consider all the particles each time step, however it also assumes that the particles were sorted by their grid index. In order to build the grid without atomic operations, a sorting implementation was used. Several kernels were written to deal with the function load of this algorithm – the initial kernel "calculateHash" computes and attributes a hash value to each particle based on its matrix entry index (ID). Even though the linear id was used, a potential optimization at this stage would be to implement Z-order curve[53] (for memory access coherence). The kernel ultimately stores the results for the particle array in global memory as a *unit2* pair (cell hash, particle id respectively).

The sorting phase on the hash values is straightforward and implements fast radix sort included in the CUDPP library[54], which returns a sorted particle id list. The second kernel of importance with the signature "identifyCellStart" is responsible for using a thread per particle in order to compare the cell index of current particle with the previous particle's index. If the two indexes differ, a new cell source has been identified, which implies that the start address will be written to another array using a scattered write. For convenience, the code implemented the reverse of this function, which identifies the cell's end in a similar manner. Depending on hardware specifications, the scattered write approach might have to be replaced by a binary search implementation, as pre-CUDA architectures lack the specific memory type to support this operation[55].

**Table 1.** Illustration of list entry based on the sorting approach

| Index | Unsorted List | | Sorted List by cell ID | Cell Start |
|---|---|---|---|---|
| | Cell ID | Particle ID | | |
| 0 | 9 | 0 | (4,3) | - |
| 1 | 6 | 1 | (4,5) | - |
| 2 | 6 | 2 | (6,1) | - |
| 3 | 4 | 3 | (6,2) | - |
| 4 | 6 | 4 | (6,4) | 0 |

The position and velocity arrays were also sorted in order to improve the efficiency of texture ID lookups during collision detection phase. Each particle will then have a grid location attributed to it which will be updated at each time-step. In addition to this, the main function of the algorithm will iterate through the neighbouring 27 adjacent matrix locations to check for any collision. If a collision is detected, in response, the velocity of the particle will be modified by an arbitrary value[56].

**Figure 6.** Capture of the uniform-grid method for the particle system.

**Growth cone steering mechanisms** – At each discrete time-step, a predefined distance is traversed by the axon's growth cone on the randomly allocated direction vector. Via the minimum and maximum detectable concentration of a diffusible guidance cue, and the neurite's range, the growth cone will either continue to traverse the environment on its current direction, or if the gradient is in range, it will cause a collision to be acknowledged and a direction change (steering) will immediately take place. The magnitude of the turning direction is given by the relation $\delta\theta° = \pi/30$ to either the left (on the x-axis) or right (this change depends on the sign on the equation). The assumption made at this stage is that the growth cone detects the change in concentration across its width, with a measurement whose SNR is related to the background concentration $\gamma$, gradient density $\mu$, gradient average direction $\theta_{NGF}$, and the growth cone's current direction dictated by:

$$SNR_{\alpha\mu}\sqrt{\frac{\gamma}{(1+\gamma)^3}}\sin(\theta - \theta_{NGF}) \quad (5)$$

The growth cone's decision to turn right or left respectively relies on a Gaussian distribution $N$ (SNR,1), where the mean is the result of the equation described above, and the variance of the distribution is one. If the sign of the distribution is positive, then the growth cone will turn to the right, otherwise, it will turn to the left. The alternative computation mechanism employed, which yielded better efficiency, but less uniform output was to compute an average direction of the particles diffusing in range of the growth cone via the vector addition method (since all particles are defined by a concentration, velocity, rate of production, and magnitude) it is a facile task to compute the resultant vector of the particles' direction and make the axon turn at each time step on the resultant vector given by adding the axon's current direction vector and the delta vector of the diffusing particles it came in range of. In order to determine the angle between the two vectors (current direction and gradient diffusion direction), a number of steps has to be taken: the components of both vectors have to be multiplied on the x-axis (in orthonormal basis):

$$v_1 = x_1 e_x + y_1 e_y + z_1 e_z \quad (6)$$
$$v_2 = x_2 e_x + y_2 e_y + z_2 e_z$$

Where by multiplying two of the same vectors $e_x e_x = 1$, the multiplication is straightforward. However, the notion of orientation comes in when multiplying two different basis vectors; "$e_x e_y$" will denote the plane spanned by $e_x$ and $e_y$ with the orientation turning $e_x$ into $e_y$ (implies that $e_x e_y = -e_y e_x$). Both the dot (scalar) and cross products are directly proportional to the magnitude of the vectors, where the magnitude (length) of a vector is given by the following formula:

$$|x| = \sqrt{x_1^2 + x_2^2 + \cdots + x_n^2}, \text{ vector x} = (x_1, x_2, x_3, \ldots, x_n). \quad (7)$$

Taking the proportion into consideration, in order to obtain the angle between the two vectors, the vectors will both have to be divided by the magnitude. In order to calculate the formula, all the above information has to be plugged in:

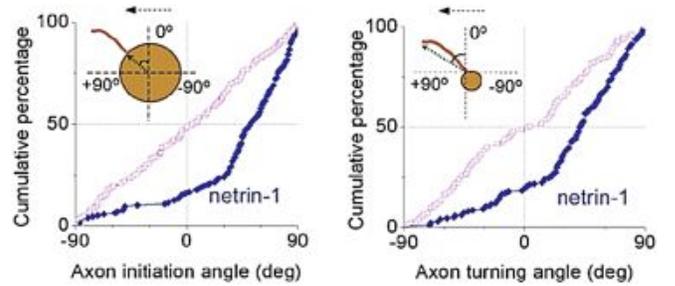

$$\sphericalangle(v_1, v_2) = \frac{v_1 v_2}{|v_1||v_2|} = \frac{(x_1 e_x + y_1 e_y)(x_2 e_x + y_2 e_y)}{\sqrt{v_1 v_1}\sqrt{v_2 v_2}}$$
$$= ([x_1 x_2 + y_1 y_2] + [x_1 y_2 - x_2 y_1]e_x e_y)/\sqrt{v_1 v_1}\sqrt{v_2 v_2}$$

At this phase of the equation, the first term located in the square brackets represents the dot product, so it is actually equivalent to the value of $\cos\theta$. Similarly, the second term represents the value of the cross product, being equivalent to $\sin\theta$.

**Figure 7.** Schematic representation of the axon's direction vector and its sequential changes depending on its detection range and overall direction of the particles it comes in contact with. After initiation, the axon should steer towards the high or low density side of the gradient respectively.

## 4 RESULTS

The following section of the paper will delineate the output of the various stages of the simulation implementation, commencing with the uniform random sphere distribution algorithms for neuron positioning and concluding with the axon guidance through cue diffusion. The results were selected from more than one set of test runs, and will reflect the areas of interest of the currently presented paper – precision (in terms of connectivity and particle interaction), efficiency (since the number of neurons that needs to be generated is large), and reusability of framework. The implementation was tested on more than one hardware set, the specifications of which are outlined below:

**Table 2.** Overview of the machines on which the simulation was tested

| Machine | CAGE038 | DAWN | Cylon |
|---|---|---|---|
| Processor | Intel® Core™ i7 2600S 2.80GHz | Intel® Xeon® CPU E5-2650L (1.80GHz) | Intel® Core™ 2 6700 (2.66GHz) |
| RAM | 8.00GB | 128GB | 2.87GB |
| OS | Windows 7 SP1 64-bit | Windows 2008 Server HPC Edition 64-bit | Windows XP Professional 32-bit |
| Graphics Card | Nvidia Quadro 600 | 2xNvidia Tesla K20 | 2xNvidia GeForce GTX 460 |
| Global Mem. | 1024MB | 5GB | 1GB |
| Global Cache | 128KB | - | 112KB |
| Local Mem. | 64KB | 64KB | 48KB |
| Max. alloc. | 1GB | 2GB | 256MB |
| Clock Speed | 1.28GHz | 2.6GHz | 1.3GHz |
| Compute Unit | 1 | 13 | 7 |
| Processing Elements | 96 | 2496 | 336 |
| Max. work group size | 3797 | - | 1024 |
| OpenGL | 4.3 | 4.3 | 3.2 |
| GLSL | 330 | 330 | 330 |

## 4.1 Initiation of neurons' position

The four algorithms used to randomly uniformly distribute points on a sphere are rejection sampling, normal deviate, spherical coordinate approach and trigonometry approach. Each one of these algorithms was tested in generating 10000 neurons, and their individual execution time was calculated. Ideally, the algorithm required for this type of simulation would output positions which are not too clustered together (the positions cannot collide, indifferent of the algorithm used) in a relatively small amount of time. Computationally, the rejection sampling and spherical coordinate approaches are efficient, however, when generating very large amounts of coordinates (100.000 or more), clustering at the poles can become imminent.

**Table 3.** Execution time vs number of neurons generated with each method

| No. neurons | Rejection Sampling | Normal Deviate | Trigonometric Approach | Coordinate Approach |
|---|---|---|---|---|
| 500 | 0.25 | 0.13 | 0.12 | 0.12 |
| 2500 | 0.40 | 0.25 | 0.24 | 0.24 |
| 5000 | 0.41 | 0.39 | 0.39 | 0.40 |
| 10000 | 0.70 | 0.69 | 0.68 | 0.70 |

The time(s) for each method was retrieved via the clock() method implemented specifically for the "GenerateCoordinate" class.

The normal-deviate method generalizes well to n-dimensions[57], however it can become computationally expensive when generating vast amounts of coordinates. In contrast, the spherical coordinate method and trigonometric method tend to commence generation closer to the sphere's origin point (O(0,0,0)), creating small clusters as they go along the distribution, which even if it is not as computationally expensive as the normal-deviate method, tend to not scale as well in n-dimensions.

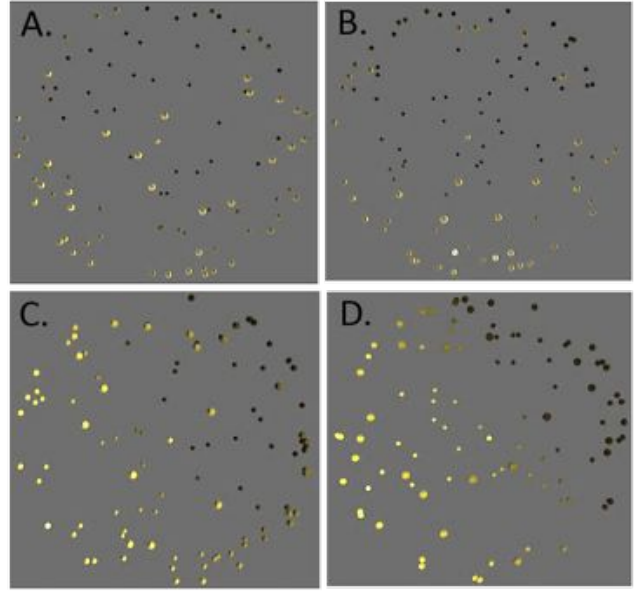

**Figure 8.** A. Rejection sampling; B. Normal-deviate; C. Trigonometric approach; D. Spherical coordinate approach.

Because the trigonometric approach does not employ rejection of generated random values, the correct coordinates are actually computed in a single loop, rendering the algorithm quite efficient, as reflected in the following plot.

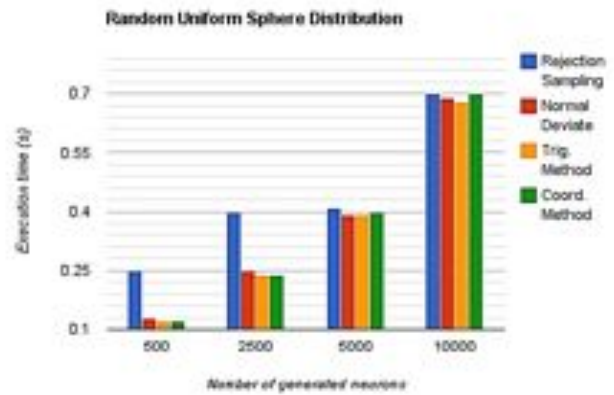

**Figure 9.** Chart illustrating the time required to generate coordinates for various numbers of neurons by each of the algorithms implemented.

## 4.2 "Direct Guidance" approach

The current approach represents a straightforward simulation of linear axon guidance, where a connection is made if the tip of the growing axon collides with another neuron (sphere-sphere collision detection method). The underlying algorithm generates the neuron positions as a Vector3f and stores this data in a list. The second important data structure employed by the current analysed approach is represented by the adjacency matrix which is computed step by step, in correlation with the collision-detection algorithm. Together, the two structures are exported to two independent spreadsheets, imported into Matlab and analysed via connectivity plots. At this phase of the simulation, the most relevant data for evaluation of efficiency and accuracy is represented by the execu-tion time of the algorithm given a certain number of steps and number of neurons, the number of connections made, and the different results obtained on different hardware sets (relevant especially for the GLSL and CUDA modules, where GPU memory manipulation plays a vital role in the overall execution time).

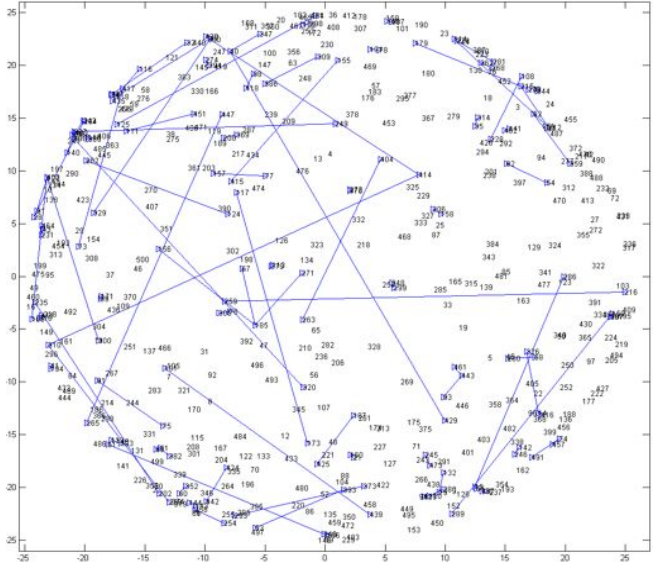

**Figure 10.** MatLab plot generated using the exported adjacency list along with a file containing a 3x500 positions representing neuron's origin.

Through the graph representation employed in the above plot, it was relatively facile to illustrate 500 neurons and the connections between each neuron, to essentially form an overall view of the network. This is constitutive information for the analysis of emerging connectivity patterns, and was achieved by exporting the adjacency matrix computed on the GPU using GLSL and CUDA along with a vector of all coordinates for the neurons into MatLab. Once the data was collected as two separate matrix data structures, it was straightforward to create a numbered sparse plot of the neuron location and overlap it with the positions in the adjacency matrix which evaluated to the value of one.

**Table 4.** Direct axon guidance results overview

| No. neurons | No. connection | Exe. Time (s) |
|---|---|---|
| 100 | 57 | 5.81 |
| 500 | 302 | 26.14 |
| 1000 | 873 | 42.67 |
| 2500 | 2013 | 105.88 (~2min) |
| 5000 | 3715 | 285.43 (~6min) |

The simulation step-size and offset of the neuron distribution were scaled, in which case for 100 neurons, the offset was of 2.5 units, and the simulation step-size was of 0.35 units (considering a total of 100 steps when the rejection sampling method was employed). Considering that the neuron distribution was in a given range *[-x, x]*, if an offset was not applies, the simulaton would have been difficult to visualise, and it would allow for more connections to be created if the number of steps was quite large. In order to maintain a realistic approach, the offset was introduced in order to allow axons to travel a minimum distance, and the total number of steps was arbitrarily chosen in order to allow any axon to cover the distance equivalent of the diameter of the distribution. The set of results presented above was obtained after running the simulation on the CAGE machine.

### 4.3 "Diffusion Guidance" approach

The model presenting axon guidance through cue diffusion implements an extra step to the direct guidance model – collision against the diffusing particles.

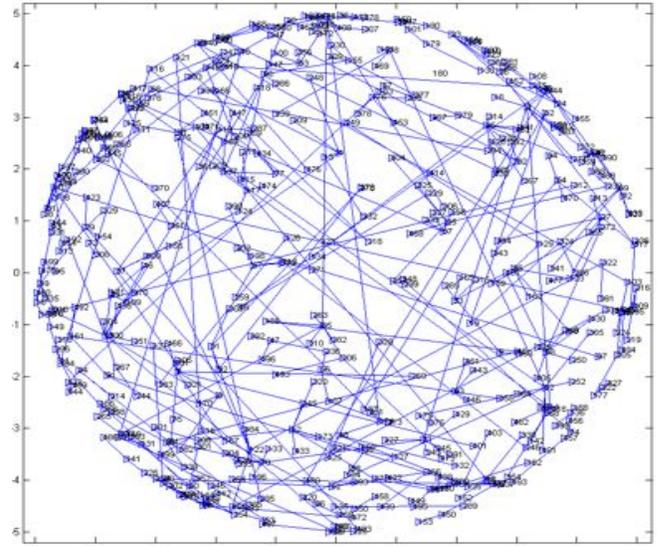

**Figure 11.** MatLab plot illustrating a connectivity graph for 500 neurons.

The diffusing cue was represented computationally by a particle system data structure, holding spherical textured particles which have various lifetimes and velocities. An arbitrary number of particle systems was instantiated to represent different cues (of various overall diffusion rates), then placed pseudo-randomly inside the neuron distribution. The simulation ran would illustrate how at each time-step the axons grow on their specific initial direction, and how this direction vector changes once the axon is in close proximity of a diffusing cue. At cue collision, the axon will steer either to the left or to the right of the diffusing cue (which means the cue attracted or repelled the motile structure of the axon's growth cone).

**Table 5.** Axon guidance through cue diffusion results overview

| No. neurons | No. connections | Execution time |
|---|---|---|
| 100 | 66 | 30.18s |
| 500 | 395 | 226.02s |
| 1000 | 778 | 300.5s |
| 2500 | | |
| 5000 | | |

The results present in the table were obtained after testing the simulation on the CAGE038 machine, with a time-step size of 0.35 units and an offset value of 2.5 units for each group of neurons tested.

In addition to this, due to the diffusing cues being present, it is highly likely that more connections would be formed through the help of the diffusing cue, so it is a good idea to scale the model to the right size, where neurons do not cluster at the poles, and where the axon travels a minimum distance, $d \geq r_{distribution}$.

As the diffusing cues represented by individual particle systems employ a sorting algorithm, the overall memory coherency at collision time is improved, and in addition to that it reduces warp divergence (particles in the same region of the simulation environment tend to have similar numbers of neighbours). Performance at this stage in the algorithm was achieved by binding global memory arrays to textures and using a texture lookup (*tex1Dfetch*), as texture reads are cached.

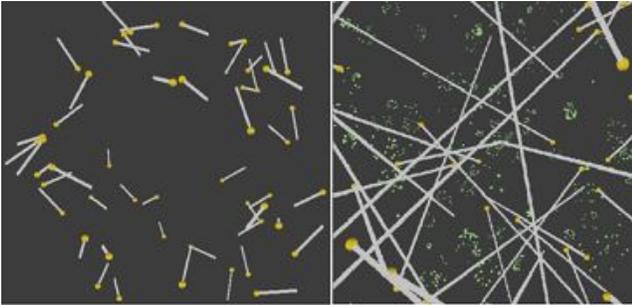

**Figure 12.** Direct axon guidance model (left hand side), and axon guidance through cue diffusion (right hand side).

## 5 DISCUSSION

### 5.1 Uniform-grid and pseudo-number generation approaches for enhancing neuron positioning

The model developed and described in this paper illustrates that computational models which make use of GPU and parallel processing power are highly suitable for simulating and analysing vast amounts of biological data accurately and efficiently, and therefore qualify as one of the most appropriate methods for simulating axon growth and guidance through cue diffusion.

The current approach presented in the paper employs four different types of algorithms for neuron distribution which eliminate the traditional grid positioning in order to avoid collision check in the primal phase of neuron coordinate generation. The algorithms developed generate positions following the mathematical model of random uniform sphere distribution which means that no two set of coordinates will collide and that the distribution takes the shape of a sphere (suitable for simulating a model related to the development of the nervous system). Even though a grid structure becomes cumbersome to traverse and would require multiple traversals if it was used in the initial process in order to eliminate identical coordinates, it still represents the best approach for splitting collision detection process into areas for the objects in the simulation. The time-efficiency gained by using mathematical approaches to generate dissimilar coordinate sets in a certain shape is lost at the second stage of the algorithm which checks for collision detection between the axon's motile growth cone and another neuron. During the collision-detection phase, if the neuron's location was stored in the grid, the area of collision could be narrowed down to the objects around the target, essentially meaning that the number of objects the algorithm would test would be halved. Ideally, after generating the neurons using one of the proposed algorithms, the neurons would be sorted and stored in a grid data structure entirely computed on the GPU using CUDA. This approach would increase efficiency at the cost of narrowing down the hardware compatibility. The four algorithms for generating coordinates could benefit from GPU computational power, as illustrated in C. L. Phillips'[58] paper regarding generation of pseudo-random values (PRNG) for dissipative particle dynamics via massively parallel processing. The paper presents an alternative to single-instruction-multiple-thread (SIMT) approach to number generation. The alternative constitutes of a one-PRNG-per-kernel-call-per-thread implementation, through which a micro-stream of values is generated in every thread and kernel call. The generated streams of values do not require global memory for state storage (no inter-thread communication is necessary), which renders the method highly efficient in comparison to the memory-bound kernel approach. Similar approaches were used for generating a bank of random values which are written to the global shared memory or for a large number of computational steps to be executed through a single kernel call[59]. However, there are trade-offs and drawbacks to this approach, such as the fact that the stream of random numbers is continuously computed over a myriad of short kernel calls, which translates into loading thread states from memory when the kernel call commences and storing the state to memory at the end of the kernel call, which is not an adequate approach since per-thread resources can be exhausted when hosting large vectors to store the generated streams of values. Generally, when employing this method the state size is approximately 35 long integers (64-bits) or 280 bytes per-thread, obliging expensive repeated reads and writes to the global memory or using minuscule thread blocks[60]. Another drawback is represented by the partitioning step which is permanently dependent on the PRNG's computational efficiency or requires a separate kernel call in order to periodically update the random number generator and cache the values in global memory for later access. This type of caching specific for random numbers relies on a data management framework inside the kernel for constantly updating the cache. Even if it is assumed that the pre-generation process of the cache came to no computational cost, the process of loading a single random value into an arbitrary thread would require more time[61] than it would take to generate it inside the specific thread from the vector state information.

### 5.2 Diffusing cue simulation via Lagrangian integration and the benefits of level-set methods

The axon guidance via cue diffusion simulation implements an Eulerian (grid-based) particle system, which involves calculating a set of particle properties at an arbitrary fixed point in space. This method could be replaced by a Lagrangian particle system, which tracks and calculates the properties of particles as they are traversing the simulation's environment. The former method is more suitable (in terms of accuracy) for the axon guidance model, and it has several advantages compared to the grid method – it only performs calculations when necessary, and therefore the bandwidth allocation is lower than the one required by a grid structure. The properties of each particle will be stored at particle position and not at every fixed point. Another very important aspect of Lagrangian integration applied to particle systems is that the particles will not be constrained to a finite box, which means that when simulating neural development, particles can travel and influence a number of areas without constraints. The reason why a grid approach was implemented in this simulation and not a Lagrangian particle approach is because in order for the former to function, a large number of particles have to be initialized as input. The optimal technique for this simulation would be represented by the particle level-set method, which combines the strengths of both a uniform grid and Lagrangian integration[62]. Generally, the level set method (LSM) is responsible for tracking topology changes of a mesh or shape surface. Since all numerical computations involving curves or surfaces can be carried out on a fixed Cartesian grid without the need to parameterize the objects in the environment, the LSM could be an ideal candidate for the particle systems required in this simulation. However, for the particle-particle interaction from within the diffusing cue a Stochastic Eulerian-Lagrangian method (SELM) should be used in order to model viscosity, electric charge and rate of diffusion over time. This method is commonly employed to capture indispensable components of fluid structure interactions subject to thermal variations. The SELM method makes extensive use of the Eulerian integration to express the hydrodynamic fields[63] in conjunction with the Lagrangian integration to describe elastic structures. An Eulerian-Lagrangian method is essentially described by a set of three equations which correlate the Eulerian expression to the Lagrangian degree of freedom.

$$\rho \frac{du}{dt} = \mu \Delta u - \nabla p + \Lambda[\Upsilon(V - \Gamma u)] + \lambda + f_{thm}(x,t)$$
$$m \frac{dV}{dt} = -\Upsilon(V - \Gamma_u) - \nabla \Phi[X] + \xi + F_{thm}$$
$$\frac{dX}{dt} = V$$

The pressure denoted by the parameter $\rho$ is characterized by the condition of incompressibility of the system $\nabla u = 0$, and $\{X, V\}$, which represent the composite vectors of the Lagrangian integration method[64,65]. The first two equations use stochastic driving fields in order to express thermal fluctuations ($f_{thm}$) and Lagrange multipliers which impose constraints such as local rigid bodies deformations. For the second equation, it is necessary to introduce the potential energy in order to configure the structure ($\Phi$). Approximations were employed in order to eliminate dynamics on small time-scales degrees of freedom.

### 5.3 Alternative approach for cue diffusion collision detection against axon's motile growth cone

In the case of axon guidance through diffusing molecules, the approach presented in this paper applied an overall detection range for each axon through which it would determine if it was or not in the close proximity of the diffusing cue. If the growth cone structure was indeed in range of the active diffusing cue, the axon would then decide to steer either left or right (based on the overall resultant direction vector of the particles which were still alive and specific to the currently evaluated diffusing cue). This renders a straightforward, relatively inexpensive computational model which simulates the desired behavior. The steering process is smooth as the direction vector is changed gradually, and the diffusion of molecules continues. However, the outcome of the simulation in terms of connectivity can be radically influenced at the sensing process level (when the growth cone is in range of the cue and detects the diffusion and gradient fluctuation). As particles are diffused and travel a certain distance away from the diffusing source through the environment, their velocity changes and becomes influenced by the environment's viscosity. Therefore, an alternative implementation would be to instantiate a Particle class which would store data relevant to the simulation's environment, such as distance travelled from source, current location (implying that the uniform-grid approach is still used), velocity, lifetime, minimum and maximum concentrations detectable. At each time step, each particle can potentially influence the axon's path, as each particle will have a direct impact on the axon's velocity of traversing that specific region in the simulation. The major differences between this implementation and the one employed in the simulation are related to computational efficiency. If each individual particle that the axon's growth cone detects would influence the axon's path, the number of connections established would increase. In addition to this, a larger number of parameters would have to be stored, compared at each step for every particle in the simulation against the axon of each unconnected neuron. This will drastically influence not only the outcome of the simulation in terms of connection density, but it will also slow the efficiency, since there would be more inter-thread activity, and memory writing and reading. However, this aspect of the simulation should be explored in the future, as it presents extensive methods through which the environment could be described better and accurate simulations could be created with a higher level of detail in terms of molecular activity.

### 5.4 Axon branching implementation for direct axon guidance and guidance through cue diffusion

During the development of the nervous system, a number of axons will end the growing process once their termination point was reached, however, it is not uncommon for axons to continue projecting past their target, loop back or even extend collateral branches. Generally they can be situated in one of the three states: growing phase, termination, or steering. The two approaches to axon guidance presented in this paper do not account for the axon branching phase, however it can be easily implemented using the currently described framework. Once an axon has reached the terminal target, if cues are in its range of detection and present a concentration which is detectable by the axon's growth cone, then the axon can continue projection, being guided along the gradient. This approach could potentially yield a larger number of connections compared to the values presented in the results section, however this will come at an efficiency cost. The number of discrete steps will have to be increased for collision detection, which directly influences the overall simulation time. On hardware with high GPU specifications presenting CUDA compatibility, the efficiency, accuracy and frame rate will remain mostly similar. However, for developing a more robust simulation, the current computational framework leaves room for implementing the axon branching feature due to the framework's object oriented design.

### 5.5 Conclusion

### ACKNOWLEDGEMENTS

The current project was developed under the careful supervision and guidance of Dr. Marcus Kaiser(Newcastle University, School of Computing Science and Institute of Neuroscience), who has pointed me in the right direction every time I have required clarifications of the underlying biophysical framework sustaining my computational model. It was a great experience to develop scientific software which brings together the complex field of neuroscience and powerful GPU computing. I also acknowledge Dr. William Blewitt for his advice on establishing a remote testing and debugging framework for MVS environment via Nvidia Nsight Monitor on a graphic server, Dr. Sol Lim for providing depth clarifications of the mathematical framework of the direct axon guidance model, and my family and fellow colleagues from the department of Bioinformatics who have offered their constructive criticism and advice on my research project.